\documentclass[aps,prd,reprint,amsmath,amssymb,superscriptaddress,showpacs,floatfix,10pt]{revtex4-1}
\usepackage{graphicx}
\usepackage{dcolumn}
\usepackage{bm}
\usepackage{hyperref}
\usepackage[usenames, dvipsnames]{color}

\begin{document}

\title{Path integral action of a particle in $\kappa$-Minkowski spacetime}

\author{Ravikant Verma}
\email{ravikant.verma@bose.res.in, ravikant.uohyd@gmail.com}
\affiliation{ Department of Theoretical Sciences, \\ S.N. Bose National Centre for Basic Sciences, Block - JD, Sector - III, Salt Lake, Kolkata - 700106, India}

\author{Debabrata Ghorai}
\email{debabrataghorai@bose.res.in}
\affiliation{ Department of Theoretical Sciences, \\ S. N. Bose National Centre for Basic Sciences, Block - JD, Sector - III, Salt Lake, Kolkata - 700106, India}
 
\author{Sunandan Gangopadhyay}
\email{sunandan.gangopadhyay@bose.res.in,\\sunandan.gangopadhyay@gmail.com}
\affiliation{ Department of Theoretical Sciences, \\  S. N. Bose National Centre for Basic Sciences, Block - JD, Sector - III, Salt Lake, Kolkata - 700106, India}


\begin{abstract}
\noindent In this letter, we derive the path integral action of a particle in $\kappa$-Minkowski spacetime. The equation of motion for an arbitrary potential due to the $\kappa$-deformation of the Minkowski spacetime is then obtained. The action contains a dissipative term which owes its origin to the $\kappa$-Minkowski deformation parameter $a$. We take the example of the harmonic oscillator and obtain the frequency of oscillations in the path integral approach as well as operator approach upto the first order in the deformation parameter $a$. For studying this, we start with the $\kappa$-deformed dispersion relation which is invariant under the undeformed $\kappa$-Poincar$\acute{e}$ algebra and take the non-relativistic limit of the $\kappa$-deformed dispersion relation to find the Hamiltonian. The propagator for the free particle in the $\kappa$-Minkowski spacetime is also computed explicitly. In the limit, $a\rightarrow 0$, the commutative results are recovered.  
\end{abstract}

\maketitle


The theory of general relativity describes the spacetime structure at the classical level and presents the notion of continuum spacetime. In contrast, quantum mechanics provides the behaviour of the physical system at the microscopic level, that is at very small length scales. These two profound theories of physics, which has changed the landscape of theoretical physics, have been very difficult to combine into a single theory. Attempts have been made in this direction to develop a unified theory and such theories are known as the quantum theories of gravity. Development of a consistent theory of quantum gravity is now one of the most active area of theoretical physics. 

Noncommutative geometry provides a possible way to capture the spacetime structure at the quantum level. The noncommutativity of spacetime was first introduced by Snyder in 1947 \cite{sny}, with the aim of handling the UV divergences in quantum field theories.  In this approach, coordinates of spacetime get quantized. This in turn leads to the modification of the symmetry of spacetime. One such deformed symmetry algebra that has attracted wide attention in the study of quantum gravity is the $\kappa$-Poincar$\acute{e}$ algebra and the associated spacetime is known as the $\kappa$-Minkowski spacetime \cite{kap2}-\cite{kap5}. This spacetime has emerged in the low energy limit of quantum gravity models \cite{A} and is one of the examples of noncommutative spacetimes where the coordinates obey the following commutation relations
\begin{equation}
[\hat{x}_i, \hat{x}_j]=0,~~ [\hat{x}_0, \hat{x}_i]=i a \hat{x}_i,~~  a=\frac{1}{\kappa}; ~~i,j= 1,2,...,n-1.
 \label{a1}
\end{equation}
where $a$ is the deformation parameter and has the dimension of length. The above algebra can be written in a covariant form in terms of the Minkowski metric $\eta_{\mu\nu} = diag(-1,1,1,....,1) $ as 
\begin{eqnarray}
[\hat{x}_{\mu}, \hat{x}_{\nu}] = i C_{\mu\nu\lambda}\hat{x}^{\lambda} =i (a_{\mu}\eta_{\nu\lambda} - a_{\nu}\eta_{\mu\lambda}) \hat{x}^{\lambda}
\end{eqnarray}
where $a_{\mu}~ (\mu,\nu=0,1,...,n-1) $ are real and dimensionful constants with $a_0= a \equiv\frac{1}{\kappa}$ and $a_{i}=0 $ for $i=1,2,...,n-1$. 
The coordinates $\hat{x}_i$ and $\hat{x}_0$ are mapped to functions of $x_0,~x_i$ and $\partial_0,~\partial_i$ (which are coordinates of the commutative spacetime and their derivatives)\cite{nc61}-\cite{u5}. Such a realization is given by \cite{u5}
\begin{equation}
\hat{x}_\mu = x^\alpha \Phi_{\mu\alpha} (\partial).
\end{equation}
It is easy to see that coordinates obey $[\partial_{\mu} , \hat{x}_\nu ]=\Phi_{\mu\nu}(\partial)$. This realization can map the function of noncommutative spacetime to the function of the commutative spacetime. This can be seen first by defining the constant function annihilated by $\partial$ as the vacuum $|0\rangle \equiv 1$ and defining \cite{u5}
\begin{equation}
F(\hat{x}_\varphi)|0\rangle =F_{\varphi}(x)\label{xy}
\end{equation}
where the subscript $\varphi$ specify the realization we are working with.
In such class of realization, the form of the noncommutative coordinates $\hat{x}_i$ and $\hat{x}_0$ satisfying the algebra (\ref{a1}) in terms of the commutative coordinates $x_i$ and $x_0$ can be written as \cite{nc61}-\cite{u5}
\begin{eqnarray}
\hat{x}_i&=&x_i \varphi(A)\label{a2}\\
 \hat{x}_0&=&x_0 \psi(A) + i a x_i \partial_i \gamma(A)\label{a3}
\end{eqnarray}
where $A=-ia\partial_0$. 

The motivation of this letter is to carry out a path integral formulation of quantum mechanics in $\kappa$-Minkowski spacetime which has so far been missing in the literature. Such a formulation has been carried out in noncommutative quantum mechanics \cite{spallucci}-\cite{mignemi} with a canonical structure of noncommutativity between coordinates, namely, $[\hat{x}, \hat{y}]=i\theta$. Our present investigation would therefore help us compare our results with those derived in \cite{spallucci}-\cite{mignemi} and would give an insight about the intricacies of quantum mechanics formulated in $\kappa$-Minkowski spacetime. It should be noted, however, that we shall consider the non-relativistic of the $\kappa$-deformed dispersion relation in our analysis. This in turn would incorporate the effects of $\kappa$-Minkowski spacetime on the non-relativistic Hamiltonian. The reason for doing this is that a path formulation with the relativistic Hamiltonian in $\kappa$-Minkowski spacetime is quite difficult to carry out technically.

The symmetry algebra of the $\kappa$-deformed spacetime known as the $\kappa$-Poincar$\acute{e}$ algebra in a classical basis reads 
\begin{equation}
[M_{{\mu}{\nu}} , D_{\lambda}]=\eta_{{\nu}{\lambda}} D_\mu - \eta_{{\mu}{\lambda}} D_\nu ,~~ [D_\mu , D_\nu]=0\label{dirac4}
\end{equation}
\begin{equation}
[M_{{\mu}{\nu}}, M_{{\lambda}{\rho}}]=\eta_{{\mu}{\rho}} M_{{\nu}{\lambda}} + \eta_{{\nu}{\lambda}} M_{{\mu}{\rho}} - \eta_{{\nu}{\rho}} M_{{\mu}{\lambda}} - \eta_{{\mu}{\lambda}} M_{{\nu}{\rho}}.\label{dirac5}
\end{equation}
This algebra can also be put in the form
\begin{equation}
[M_{ij}, M_{j0}] = M_{i0} 
\label{eqd1}
\end{equation}
\begin{equation}
[M_{i0}, M_{j0}] = M_{ij}.
\end{equation}
The action of the generalized rotation generators $M_{\mu\nu}$ 
on the noncommuting coordinates $\hat{x}_{\mu}$ keeping the Lie algebraic structure yields \cite{gupta},\cite{u5}
\begin{equation}
[M_{i0}, \hat{x}_{0}] = -\hat{x}_{i} + ia M_{i0}
\end{equation}
\begin{equation}
[M_{i0}, \hat{x}_{j}] = -\delta_{ij}\hat{x}_0 + iaM_{ij}.
\label{eqd4}
\end{equation}
Requiring the generators $M_{\mu\nu}$ to be linear in $x$ and having an infinite series in $\partial$, leads to two set of solutions 
compatible with eq.(s)(\ref{eqd1}-\ref{eqd4}). One set is for $\psi=1$ and the other set is for $\psi= 1+2A$. In this letter, we shall consider the first set which leads to 
\begin{equation}
M_{ij}=x_i \partial_j - x_j\partial_i 
\end{equation}
\begin{equation}
M_{i0}=x_i \partial_0 \varphi \frac{e^{2A}-1}{2A}-x_0 \partial_i \frac{1}{\varphi}+iax_i \nabla \frac{1}{2\varphi}-iax_k \partial_k \partial_i \frac{\gamma}{\varphi}
\end{equation}
\begin{equation}
D_i=\partial_i \frac{e^{-A}}{\varphi},~~ D_0=\partial_0 \frac{\sinh A}{A} - i a \nabla \frac{e^{-A}}{2 \varphi^2}~. 
\end{equation}
The twisted coproduct of the generators $M_{\mu\nu}$ are given by
\begin{equation}
\bigtriangleup_\varphi (M_{ij})=M_{ij} \otimes I + I\otimes M_{ij} =\bigtriangleup_0 (M_{ij})
\end{equation}
\begin{equation}
\bigtriangleup_\varphi (M_{i0})=M_{i0}\otimes I + e^A \otimes M_{i0} + ia\partial_j \frac{1}{\varphi(A)} \otimes M_{ij}.
\end{equation}
The $\kappa$-Poincar$\acute{e}$ algebra defined in eq.(s)(\ref{dirac4}, \ref{dirac5}) have the same form as the Poincar$\acute{e}$ algebra in commutative spacetime. But the generators of the algebra get modified due to the $\kappa$-deformation of the spacetime. In this letter, 
we shall use the $\kappa$-Poincar$\acute{e}$ algebra given in eq.(s)(\ref{dirac4}, \ref{dirac5}).

The Casimir of this undeformed $\kappa$-Poincar$\acute{e}$ algebra, $D_{\mu}D^{\mu}$ reads
\begin{equation}
D_{\mu}D^{\mu}=\square (1 + \frac{a^2}{4} \square).
\end{equation}
The explicit form of the $\square$ operator is given by
\begin{equation}
\square = \nabla \frac{e^{-A}}{\varphi^2} + \frac{2\partial^2_0}{A^2}(1 - \cosh A). 
\end{equation}
In the $\kappa$-deformed Minkowski spacetime, the dispersion relation is obtained from the Casimir by
\begin{equation}
P_\mu P^\mu=P_0 P^0 + P_i P^i =-m^2,~~~P_\mu=-iD_\mu
\end{equation}
where the momenta $P_0$ and $P_i$ are the $\kappa$-deformed momenta. Explicitly, the $\kappa$-deformed dispersion relation is given as \cite{gupta}, \cite{u5}
\begin{eqnarray}
&\frac{4}{a^2}&\sinh^2 \Big(\frac{aE}{2c}\Big)-p_i^2 \frac{e^{-a\frac{E}{c}}}{\varphi^2(a\frac{E}{c})}\nonumber\\ &+& \frac{a^2}{4}\left[\frac{4}{a^2}\sinh^2\Big(\frac{aE}{2c}\Big)-p_i^2 \frac{e^{-a\frac{E}{c}}}{\varphi^2(a\frac{E}{c})} \right]^2=m^2c^2.\label{ll}
\end{eqnarray}
In the limit $a\rightarrow 0$, we get the dispersion relation in the commutative spacetime.

With the above formalism in place, we now proceed to formulate path integral in the $\kappa$-deformed spacetime. To begin with, we first take the non-relativistic limit of the $\kappa$-deformed dispersion relation given in eq.(\ref{ll}). This is done since  the path integral formulation is easier to carry out in a non-relativistic setting. After straight forward calculation for the choice of $\varphi(A)=e^{-A}$, we obtain the following form of the Hamiltonian 
\begin{equation}
H= \frac{1+acm}{2m}\vec{p} \cdot \vec{p} \equiv \frac{\vec{p}^2}{2m_{eff}}\label{uu}
\end{equation}
where 
\begin{equation}
m_{eff}=\frac{m}{1+acm}~.
\end{equation}
This is the free particle Hamiltonian upto first order in the deformation parameter $a$ and $\vec{p}$ is the momentum in the commutative spacetime. 
To proceed further, we substitute eq.(s)(\ref{a2}, \ref{a3}) in eq.(\ref{a1}) and assuming that $\varphi(A)$ and $\psi(A)$ are positive functions and satisfies $\varphi(0)=1, ~\psi(0)=1$ yields 
\begin{equation}
\frac{\varphi^{\prime}}{\varphi} \psi = \gamma -1.
\end{equation}
Now for the choice of $\varphi(A)=e^{-A}$ and $\psi(A)=1$, we get $\gamma=0$. This leads to 
\begin{equation}
\hat{x}_0 = x_0,~~~\hat{x}_i = x_i e^{-A}.\label{xy1}
\end{equation}
Using eq.(s)(\ref{xy}, \ref{xy1}), we obtain \cite{hari}
\begin{equation}
\hat{x}_i^2 = \hat{x}_i \hat{x}_i \rightarrow x_ie^{-A}x_ie^{-A} |0\rangle = x_i^2 + \frac{a}{c}x_i\dot{x}_i
\label{corr}
\end{equation}
where we have used $A=-a\partial_{0}$. In writing the above relation, we have used the fact that $\partial$ annihilates the vacuum. Alternatively, one can also obtain the relation by using \cite{u5}
\begin{equation}
(\hat{x}_{\varphi})_i f(\hat{x}_{\varphi}) |0\rangle = x_{i}\varphi(A) f(x) .
\end{equation}
It should be noted however that the correction term involving the $\kappa$-deformation parameter $a$ in eq.(\ref{corr}) arises only if one assumes that the variables that one gets after quantization (eq.(\ref{xy})) are time dependent.

\noindent From the above relation, we get
\begin{eqnarray}
\hat{x}_i &=&\sqrt{\hat{x}_i \hat{x}_i}= \left( x_i^2 + \frac{a}{c}x_i\dot{x}_i \right)^\frac{1}{2} =  x_i\left( 1+ \frac{a\dot{x}_i}{cx_i}\right)^\frac{1}{2}\nonumber\\ &\approx & \left( x_i+ \frac{ap_i}{2mc}\right).
\end{eqnarray}
For simplicity, we shall now write the Hamiltonian (\ref{uu}) in one spatial dimension. In the presence of an arbitrary potential $V(\hat{x})$, the Hamiltonian (\ref{uu}) can be written as
\begin{equation}
H= \left(\frac{1+acm}{2m}\right) p^2 + V\left( x+ \frac{ap}{2mc}\right).
\end{equation}
It should be noted that restricting to one spatial 
dimension still captures the effects of the $\kappa$-Minkowski spacetime on the non-relativistic Hamiltonian. Using Taylor series expansion, we get
\begin{equation}
H= \left(\frac{1+acm}{2m}\right) p^2 + V(x) + \frac{ap}{2mc}\frac{\partial V(x)}{\partial x}.\label{w}
\end{equation}
This is the Hamiltonian of the system upto first order in the deformation parameter $a$. Here note that $V(x)$ is the potential of the system in the commutative spacetime. We can now write the Hamiltonian for the harmonic oscillator $(V(x)=\frac{1}{2}m\omega^2x^2)$ upto first order in the deformation parameter $a$ as
\begin{eqnarray}
H &=& \left(\frac{1+acm}{2m}\right) p^2 + \frac{1}{2}m\omega^2 x^2 + \frac{ap}{2c}\omega^2 x\nonumber\\
&=& \frac{p^2}{2m_{eff}} + \frac{1}{2}m_{eff}\omega^2 \Big ( x^2 + \frac{ap}{2mc}x\Big) (1+acm).\label{qqq}
\end{eqnarray}
We shall now proceed to construct the path integral for this Hamiltonian written down in the commutative coordinates. For that we require the momentum eigenstate $|p\rangle$ satisfying
\begin{equation}
p|p\rangle = p |p\rangle,~~~~\langle p^\prime |p\rangle =\delta(p-p^\prime) .
\end{equation}
The following completeness relations are also crucial for the construction of the path integral
\begin{eqnarray}
\int dp ~|p\rangle\langle p| =1\\
\int dx ~|x,t\rangle\langle x,t| =1.
\end{eqnarray}
The kernel can be written as
\begin{eqnarray}
\langle x_{f}t_{f}|x_{i}t_{i} \rangle &=& \int ...\int dx_1 dx_2...dx_n \langle x_{f}t_{f}|x_{n}t_{n} \rangle\nonumber\\ 
&\times&\langle x_{n}t_{n}|x_{n-1}t_{n-1} \rangle...\langle x_{1}t_{1}|x_{i}t_{i} \rangle .\label{y1}
\end{eqnarray}
Now the propagator over the small segment in the path integral reads \cite{hibbs},\cite{rd}
\begin{equation}
\langle x_{j+1}t_{j+1}|x_{j}t_{j} \rangle=\langle x_{j+1}|e^{-\frac{iH\tau}{\hbar}}|x_j\rangle.
\end{equation}
Using eq.(\ref{w}) in the above equation, we get  
\begin{eqnarray}
\langle x_{j+1}t_{j+1}|x_{j}t_{j} \rangle &=& \int \frac{dp}{2\pi\hbar} \exp \Bigg\{ \frac{i}{\hbar}\Bigg [p_j(x_{j+1}-x_j)\nonumber\\
&-& \tau \bigg ( \frac{1+acm}{2m}\bigg )p_j^2 - \tau \frac{a}{2mc} \frac{\partial V}{\partial x}p_j\nonumber\\
& -& \tau V\Bigg (\frac{x_{j+1}+x_j}{2}\Bigg) \Bigg ]\Bigg \} .
\end{eqnarray}
Substituting this in eq.(\ref{y1}), we obtain
\begin{eqnarray}
\langle x_f t_f|x_i t_i\rangle &=& \lim_{n\to\infty}\left( \frac{m}{ih\tau(1+acm)}\right)^\frac{n+1}{2}\nonumber\\ &\times & \int \prod_{j=1}^{n} dx_j \exp  \Bigg \{\sum_{j=0}^n \frac{i\tau}{\hbar}\nonumber\\ &\Bigg [& \frac{m}{2(1+acm)} \Bigg (\frac{x_{j+1}-x_j}{\tau} - \frac{a}{2mc}\frac{\partial V(x)}{\partial x}\Bigg )^2\nonumber\\ &-& V\Bigg (\frac{x_{j+1}+x_j}{2}\Bigg)\Bigg]\Bigg\}.\label{jj}
\end{eqnarray}
In the limit $\tau \rightarrow 0$, the above expression for the path integral representation of the propagator can be written as
\begin{eqnarray}
\langle x_f t_f|x_i t_i\rangle = N\int {\cal{D}}x~e^{\frac{i}{\hbar} S}
\end{eqnarray}
where the action $S$ is given by 
\begin{equation}
S=\int dt~\Bigg[ \frac{1}{2}\frac{m}{1+acm} \Bigg( \dot{x} - \frac{a}{2mc}\frac{\partial V(x)}{\partial x}\Bigg)^2 -V(x)\Bigg].
\end{equation}
The above equation gives the action of a particle in the $\kappa$-Minkowski spacetime. The form of the action differs from that derived in \cite{sg} where the action for the particle was non-local in form with the non-locality owing its origin to the noncommutative parameter $\theta$ appearing in the commutation relation
\begin{equation}
[\hat{x}, \hat{y}]=i \theta .
\end{equation}
From this action, we can write the Lagrangian of the particle to be
\begin{equation}
L= \frac{1}{2}\frac{m}{1+acm} \Bigg( \dot{x} - \frac{a}{2mc}\frac{\partial V(x)}{\partial x}\Bigg)^2 -V(x)
\end{equation}
which upto first order in the deformation parameter $a$ simplifies to
\begin{equation}
L= \frac{1}{2}\frac{m}{1+acm} \Bigg( \dot{x}^2 - \frac{a\dot{x}}{mc}\frac{\partial V(x)}{\partial x}\Bigg) -V(x).
\end{equation}
The equation of motion following from this Lagrangian reads
\begin{eqnarray}
\Bigg( \frac{m}{1+acm}\Bigg)\ddot{x} &+& \frac{1}{2}\Bigg(\frac{1}{1+acm}\Bigg) \frac{a\dot{x}}{c}\frac{\partial}{\partial x}\Bigg(\frac{\partial V(x)}{\partial x}\Bigg)\nonumber\\ &+& \frac{\partial V(x)}{\partial x}=0. \label{z}
\end{eqnarray} 
It is interesting to note that a dissipative term arises in the equation of motion due to the $\kappa$-deformation of the spacetime. But this dissipative term vanishes in the commutative spacetime, that is in the limit $a\rightarrow 0$.

We now solve this equation of motion for two simple cases, namely, the free particle and the harmonic oscillator. For the free particle ($V=0$), the equation of motion is $\ddot{x}=0$ which is the same as the commutative case, that is the $\kappa$-deformation of the spacetime does not affect the equation of motion. But mass of the particle gets modified due to the $\kappa$-deformation of the spacetime.

Now we calculate the propagator for the free particle in $\kappa$-Minkowski spacetime. For this, we start from eq.(\ref{jj}) and set $V=0$ to get
\begin{eqnarray}
\langle x_f t_f|x_i t_i\rangle &=& \lim_{n\to\infty}\left( \frac{m}{ih\tau(1+acm)}\right)^\frac{n+1}{2}\nonumber\\ && \int \prod_{j=1}^{n} dx_j \exp  \Bigg \{ \frac{im}{2\hbar \tau(1+acm)}\nonumber\\ && \sum_{j=0}^n \Big (x_{j+1}-x_j \Big )^2\Bigg\}.
\end{eqnarray} 
From straight forward calculation, we obtain the propagator for the free particle in $\kappa$-deformed spacetime as
\begin{equation}
\langle x_f t_f|x_i t_i\rangle=\Bigg[ \frac{m_{eff}}{ih(t_f - t_i)}\Bigg]^\frac{1}{2}\exp \Bigg[\frac{im_{eff}}{2\hbar}\frac{(x_f - x_i)^2}{(t_f - t_i)}\Bigg].
\end{equation}
We make a few observations now. The form of the propagator is the same as in commutative spacetime \cite{hibbs} but the mass of the particle does get affected due to the $\kappa$-deformation of the spacetime. Further, we note that the free particle propagator has a simple form and differs from that obtained in the noncommutative space with a canonical structure $[\hat{x},\hat{y}] = i\theta $ between the spatial coordinates \cite{spallucci}, \cite{sg}. A possible reason for this could be the inherent difference between the noncommutative structures of the $\kappa$-Minkowski spacetime and the canonical noncommutativity between spatial coordinates. In the former case, the space and time coordinates do not commute whereas in the latter case, the spatial coordinates do not commute. We would also like to mention that the matrix elements, involved in the computation of the propagator in the $\kappa$-Minkowski spacetime, are between states in the commutative spacetime which is in contrast to those in \cite{spallucci},\cite{sg}. This is done to simplify the analysis. We shall see in the subsequent discussion that the results obtained from this approach agree with those obtained from the operator approach upto first order in the deformation parameter $a$. 

For the harmonic oscillator, $V=\frac{1}{2}m\omega^2 x^2$. Using this potential in eq.(\ref{z}), we find the equation of motion to be
\begin{equation}
\ddot{x} + \frac{2\omega^2}{2c}\dot{x} + (1+acm)\omega^2 x=0.
\end{equation}
The solution of the above equation reads
\begin{equation}
x(t) = B e^{-\gamma t}\sin (\omega_1 t + \phi)
\end{equation}
where
\begin{equation}
\gamma = \frac{a\omega^2}{4c}, ~~~\omega_1 =\omega\sqrt{1+acm - \frac{a^2 \omega^2}{16c^2}}
\end{equation}
and $B$ is an arbitrary constant and 
\begin{equation}
\omega_1=\Big(1+\frac{acm}{2}\Big)\omega
\end{equation}
is the modified frequency due to the $\kappa$-deformation of the spacetime upto first order in the deformation parameter $a$. \\
We now proceed to obtain the frequency using the operator approach. The Hamiltonian (\ref{qqq}) can be written as
\begin{equation}
H = \frac{p^2}{2m_{eff}} + \frac{1}{2}m_{eff}\omega^2 X^{\prime 2}. \label{t}
\end{equation}
Here 
\begin{equation}
X^{\prime 2}=x^2 +\frac{ap}{2mc}x +acmx^2
\end{equation}
which in turn yields
\begin{equation}
X^{\prime}=x +\frac{ap}{4mc} +\frac{acm}{2}x
\end{equation}
upto first order in the deformation parameter $a$. We can now rewrite eq.(\ref{t}) as
\begin{equation}
H = \frac{\hbar\omega}{2}(p^{\prime 2} + q^{\prime 2})\label{tt}
\end{equation}
where $p^\prime = \frac{p}{\sqrt{\hbar m_{eff}\omega}}$ and $q^\prime = X^\prime \sqrt{\frac{m_{eff}\omega}{\hbar}} $. From these, we have the following commutation relations
\begin{equation}
[X^\prime,~p]=i\Big (1+\frac{acm}{2}\Big), ~~~~[q^\prime,~p^\prime]=i\Big (1+\frac{acm}{2}\Big).
\end{equation} 
Introducing the annihilation and creation operators as
\begin{equation}
\hat{a}=\frac{1}{\sqrt{2\Big(1+\frac{acm}{2}\Big)}}(q^\prime +ip^\prime)
\end{equation}
\begin{equation}
\hat{a}^\dagger=\frac{1}{\sqrt{2\Big(1+\frac{acm}{2}\Big)}}(q^\prime -ip^\prime)
\end{equation}
and noting that $[\hat{a},~\hat{a}^\dagger]=1$, we find the number operator $\hat{a}^\dagger \hat{a}$ to be
\begin{equation}
\hat{a}^\dagger \hat{a} = \frac{p^{\prime 2} + q^{\prime 2}}{2\Big(1+\frac{acm}{2}\Big)} - \frac{1}{2}.
\end{equation}
Using this in eq.(\ref{tt}), we finally obtain
\begin{eqnarray}
H &=& \hbar \omega \Big (1+\frac{acm}{2}\Big) \Big[\hat{a}^\dagger \hat{a}+\frac{1}{2}\Big]\nonumber\\
&\equiv & \hbar \omega^\prime \Big[\hat{a}^\dagger \hat{a}+\frac{1}{2}\Big]
\end{eqnarray}
where
\begin{equation}
\omega^\prime= \omega \Big (1+\frac{acm}{2}\Big) ~.
\end{equation}
This matches exactly with the result that we have obtained from the path integral approach upto first order in the deformation parameter $a$.

In this letter, we have carried out the path integral formulation of quantum mechanics in the $\kappa$-Minkowski spacetime. 
Starting from the $\kappa$-deformed relativitic dispersion relation, we have first taken the non-relativistic limit of this to obtain the $\kappa$-deformed non-relativitic Hamiltonian upto first order in the $\kappa$-deformation parameter $a$. With this Hamiltonian, we have then obtained 
the action of a non-relativistic particle moving in one spatial dimension for the generic potential in $\kappa$-Minkowski spacetime from the path integral representation of the propagator. From the action, we next obtained the equation of motion for the particle and observed that there is a dissipative term present in the equation of motion which owes its origin to the $\kappa$-deformed spacetime. This is one of the main findings in this letter. This term is absent in normal quantum mechanics. After that we studied the case of the free particle and the harmonic oscillator. In the free particle case, we computed the path integral propagator in the $\kappa$-deformed spacetime and noticed that only the mass of the free particle gets affected due to the $\kappa$-deformation of the spacetime. The free particle propagator also has a simple form and differs from the result found in the noncommutative space with a canonical structure $[\hat{x},\hat{y}] = i\theta $ between the spatial coordinates. We also mentioned a possible reason for this owing to the inherent difference between the noncommutative structures of the $\kappa$-Minkowski spacetime and the canonical noncommutativity between spatial coordinates.
In case of the harmonic oscillator, we obtained the frequency and found that it gets modified due to the $\kappa$-deformation of the spacetime. The result turns out to be the same as the $\kappa$-deformed frequency of the harmonic oscillator upto first order in the $\kappa$-deformation parameter $a$ computed using the operator approach.\\

\section*{Acknowledgments}
DG would like to thank DST-INSPIRE for  financial  support. S.G. acknowledges  the  support  by  DST  SERB  under  Start  Up
Research Grant (Young Scientist), File No.YSS/2014/000180. SG also acknowledges the support under the Visiting Associateship programme of IUCAA, Pune. The author would like to thank the referees for very useful comments. \\

\textbf{Author contribution statement }:
All authors have contributed equally.


\begin{thebibliography}{99} 

\bibitem{sny}H.S. Snyder, ``Quantized Space-Time", Quantized Space-Time, Phys. Rev. 71 (1947) 38.
\bibitem{kap2}J. Lukierski, A. Nowicki, H. Ruegg, ``New quantum Poincar$\acute{e}$ algebra and $\kappa$-deformed field theory", Phys. Lett. B 293 (1992) 344.
\bibitem{kap3}J. Lukierski, H. Ruegg, ``Quantum $\kappa$-Poincar$\acute{e}$ in any Dimensions", Phys. Lett. B 329 (1994) 189.
\bibitem{kap5}J. Lukierski, H. Ruegg, W. J. Zakrzewski, ``Classical and Quantum Mechanics of Free  Relativistic Systems", Annals Phys. 243  (1995) 90.
\bibitem{A}G. Amelino-Camelia, L. Smolin, A. Starodubtsev, ``Quantum symmetry, the cosmological constant and Planck scale phenomenology",  Class. Quant. Grav. 21 (2004) 3095.
\bibitem{nc61}S. Meljanac, M. Stojic, ``New realizations of Lie algebra $\kappa$-deformed Euclidean space", Eur. Phys. J. C 47 (2006) 531.
\bibitem{gupta} S. Meljanac, A.Samsarov, M.Stojic, K.S. Gupta, ``$\kappa$-Minkowski spacetime and the star product realizations", Eur. Phys. J C 53 (2008) 295.
\bibitem{u5}T.R. Govindarajan, K. S. Gupta, E. Harikumar, S. Meljanac, D. Meljanac, ``Deformed Oscillator Algebras and QFT in kappa-Minkowski Spacetime", Phys. Rev. D80 (2009) 025014. 
\bibitem{spallucci} A. Smailagic, E. Spallucci, ``Feynman path integral on the non-commutative plane", J. Phys. A: Math. Gen. 36 (2003) L467.
\bibitem{sg} S. Gangopadhyay, F.G. Scholtz, ``Path-Integral Action of a Particle in the Noncommutative Plane", Phys. Rev. Lett. 102 (2009) 241602. 
\bibitem{sg1} S. Gangopadhyay, F.G. Scholtz, ``Path integral action of a particle in a magnetic field in the noncommutative plane and the Aharonov-Bohm effect", J.Phys. A47 (2014) 075301.
\bibitem{mignemi} S. Mignemi, R. Strajn, ``Path integral in Snyder space", Phys. Lett. A 380 (2016) 1714.
\bibitem{hari} E. Harikumar, A. K. Kapoor, ``Newton's Equation on the kappa space-time and the Kepler problem", Mod.Phys.Lett. A25 (2010) 2991.
\bibitem{hibbs}R. P. Feynman, A. R. Hibbs, ``Quantum mechanics and path integrals", Dover Publications.
\bibitem{rd} L.H. Ryder, ``Quantum Field Theory",  Cambridge University Press; 2 edition (June 13, 1996).
\end{thebibliography}
\end{document}